\documentclass{jfm}
\usepackage{graphicx,bm}
\usepackage{epstopdf, epsfig}
\usepackage{amsmath}

\usepackage[normalem]{ulem}
\usepackage{color}

\shorttitle{Edge tracking in spatially developing boundary layer flows}
\shortauthor{M. Beneitez, Y. Duguet, P. Schlatter and D. S. Henningson}

\title{Edge tracking in spatially developing boundary layer flows}

\author{Miguel Beneitez\aff{1}
  \corresp{\email{beneitez@mech.kth.se}},
  Yohann Duguet\aff{2}, Philipp Schlatter\aff{1}
 \and Dan S. Henningson\aff{1}}

\affiliation{ \aff{1}Linn\'e FLOW Centre and Swedish e-Science Research Centre (SeRC), KTH Mechanics, Royal Institute of Technology, SE-100 44 Stockholm, Sweden
\aff{2}LIMSI-CNRS, UPR 3251, Universit\'e Paris-Saclay, F-91403, Orsay, France}

\begin{document}

\maketitle

\begin{abstract}
%\highlight{PS: suggestion how to work:} use \rev{for new stuff}, \revd{to remove} and \revdd{to replace}{with new}. And of course \highlight{PS: for a comment, use your initials.}

Recent progress in understanding subcritical transition to turbulence is based on the concept of the edge, the manifold separating the basins of attraction of the laminar and the turbulent state. Originally developed in numerical studies of parallel shear flows with a linearly stable base flow, this concept is adapted here to the case of a spatially developing Blasius boundary layer. Longer time horizons fundamentally change the nature of the problem due to the loss of stability of the base flow due to Tollmien--Schlichting (TS) waves. We demonstrate, using a moving box technique, that efficient long-time tracking of edge trajectories is possible for the parameter range relevant to bypass transition, even if the asymptotic state itself remains out of reach. The flow along the edge trajectory features streak switching observed for the first time in the Blasius boundary layer. At long enough times, TS waves co-exist with the coherent structure characteristic of edge trajectories. In this situation we suggest a reinterpretation of the edge as a manifold dividing the state space between the two main types of boundary layer transition, \textit{i.e.}\ bypass transition and classical transition.
\end{abstract}


\section{Introduction}

%INTRO ON BOUNDARY LAYER TRANSITION !

%Classical scenarios
%\cite{schlichting2016boundary,schmid2012stability, boiko1994experiments}
%Bypass transition
%\cite{morkovin1969many,klebanoff1955characteristics,kline1967structure,matsubara2001disturbance,brandt2004transition}\\

%Related mechanisms for finite-amplitude streaks
%\cite{kachanov1994physical,henningson1993mechanism,andersson2001breakdown,vaughan2011stability,hack2014streak}\\

%Tools introduced for subcritical transition
%\cite{itano2001dynamics,skufca2006edge,schneider2006edge,duguet2009localized}, \cite{eckhardt2007turbulence}\\

%Adaptation to edge tracking in BL for short to moderate times+findings
%\cite{cherubini2011edge,duguet2012self, biau2012laminar}\\

%Generalisation to parallel BL flows, strengthen interest for edge tracking
%\cite{biau2012laminar,khapko2016edgeedge}
%\cite{kreilos2016bypass,khapko2016edge}\\

%Present motivation
%\cite{zammert2017harbingers,zammert2017transition}\\

%Streak/TS wave interaction
%\cite{zammert2017transition},\cite{cossu2004tollmien,xu2017destabilisation}\\

Understanding the onset of transition in boundary layer flows has always been an important challenge in aerodynamics
because of the high drag associated with turbulent flows. It is crucially dependent on the level of ambient turbulence
present in the free stream, both for aeronautic applications and in wind tunnels. An archetype for the investigation of boundary layer transition is the incompressible Blasius boundary layer. The corresponding laminar base flow is two-dimensional and develops spatially with the distance downstream. Two main transition scenarios dominate this flow case depending on the incoming level of perturbations. For weak incoming levels corresponding to flight mode in a calm atmosphere, the classical scenario predicts linear instability of the base flow to two-dimensional, spanwise-invariant Tollmien--Schlichting (TS) waves, followed by their further destabilisation \citep{SchubauerSkramstad1948,boiko1994experiments, schmid2012stability, schlichting2016boundary}. For higher turbulence levels (typically for perturbation velocities exceeding 2$\%$ of the free-stream velocity) akin to those found in turbomachines, the so-called bypass route starts at earlier streamwise locations. It features streamwise velocity defects (“streaks”) sustained by streamwise vorticity via the lift-up mechanism, and growing rapidly downstream \citep{klebanoff1955characteristics,kline1967structure,morkovin1969many}. Those streaks support further instabilities \citep{henningson1993mechanism,kachanov1994physical,andersson2001breakdown,jacobsdurbin2001,brandt2004transition,schlatter2008,vaughan2011stability,hack2014streak} leading to the nucleation of turbulent spots that invade the flow \citep{matsubara2001disturbance,brandt2004transition,kreilos2016bypass}. For sufficiently perturbed inflow, no TS wave has been convincingly reported in the bypass picture. 

Linear stability of the Blasius boundary layer profile can be carried out relatively easily by freezing the base flow and parametrising it by a Reynolds number $Re_{\delta^{*}}=U_{\infty}\delta^{*}/\nu$, where $\delta^{*}$ the displacement thickness for the laminar base flow, $U_{\infty}$ is the free-stream velocity, and $\nu$ the kinematic viscosity of the fluid. This Reynolds number is understood here as a parameter whereas for genuine spatially developing flows it increases with the distance downstream and should rather be interpreted as a streamwise coordinate. The onset of TS waves is reported from linear stability analysis at $Re_{\delta^{*}} \approx 520$ \citep{jordison1970}, with moderate corrections when taking non-parallelism into account \citep{BerlinPhD}. However bypass transition has been reported for finite-amplitude disturbances introduced as early as $Re_{\delta^{*}} \approx 300$ \citep{jacobsdurbin2001}. This suggests that bypass transition falls \emph{a priori} into the category of subcritical flows such as pipe flow, channel flow, plane Couette flow or the asymptotic suction boundary layer (ASBL), for which turbulence can exist despite the linear stability of the base flow \citep{eckhardt2007turbulence}. Borrowing concepts and toolboxes from subcritical transition appears hence as promising. In particular, a useful and recent nonlinear concept developed in wall-bounded flows is the \emph{edge manifold}, the state space boundary that separates the respective basins of attraction of the laminar and the turbulent state (assumed here an attractor). This manifold $\Sigma$ is formally of codimension one, and is the stable manifold of a simpler low-dimensional invariant regime called the \emph{edge state}. In principle all trajectories starting on $\Sigma$ reach asymptotically this edge state, while any small perturbation to such an \emph{edge trajectory} is on the verge of both relaminarising and transitioning \citep{schneider2006}. This property makes edge states, together with the set of all edge trajectories, the alternative nonlinear base flow of interest to predict bypass transition and therefore to prevent or to control it. Besides, edge states and their approximations display robust localisation in physical space as soon as the numerical domain allows for it \citep{duguet2009localized,khapko2016edge}. Recently, supported by numerical evidence that even arbitrary initial noise allows for a transient detection of the edge regime \citep{khapko2016edge}, the concept of edge state and its instability have been used to model the nucleation process of turbulent spots in the presence of strong free-stream turbulence \citep{kreilos2016bypass}. It is thus essential for the understanding of spot nucleation to gather additional knowledge about edge states in spatially developing boundary layer flows.

Former attempts at identifying edge states in boundary layer flows fall into two categories. The first category relies on different parallel approximations of the Blasius profile \citep{biau2012laminar,khapko2013localized,khapko2014complexity,khapko2016edge}. Robust slow dynamics emerge in all cases: low- and high-speed streaks burst quasi-periodically before they switch location. The second category is free from the parallel assumption and gathers edge computations in  longer domains \citep{cherubini2011edge,duguet2012self}, although with some limitations, being the domain size or spanwise symmetry. The coherent structure emerging in all cases consists of an elongated localised pair of streaks, whose three spatial dimensions grow self-similarly with the variable $\delta^{*}$. Being computationally much more demanding, this approach shows severe limitations in time. The longest bisection to date, originally performed with imposed spanwise symmetry, suggests that the symmetric edge trajectory also features quasi-cyclic regeneration of the low-speed streak in terms of rescaled time variables. Importantly, none of these investigations has reported the presence of TS waves, though such waves are expected to grow exponentially in time provided $Re_\delta^{*}$ can get large enough. 

The present study aims at revisiting edge computations in the Blasius boundary layer, by extending their time horizon sufficiently far that the nonlinear interaction between TS waves and the streaks manifests itself. An earlier study of the interaction between streaks and TS waves by superimposing both disturbances can be found in \cite{schlatter_2010}. In our case, the interaction is revealed by combining the classical bisection method with a numerical technique tailored to deal with moving localised disturbances. As we shall see, this calls for a generalisation of the concept of edge state to cases where the governing base flow instability is supercritical rather than subcritical. This situation is not specific to the Blasius profile and is expected in other shear flows such as channel flow, which also loses its stability at a finite value of the Reynolds number \citep{zammert2017harbingers, zammert2017transition}.

The structure of the paper is at follows. \S 2 is devoted to the numerical aspects of this work, whereas the main results are given in \S 3. Eventually, all the results and their implications are discussed in \S 4.

\section{Computational methodology}

\subsection{Flow set-up}

The Blasius boundary layer is the incompressible boundary layer flow over a flat plate with zero pressure gradient. Let $x,y,z$ denote respectively the streamwise, wall-normal and spanwise coordinates, $x$ being measured from the inlet of the domain, itself located at a distance $x_0$ from the leading edge of the plate. Let  $\textbf{v}=(v_x,v_y,v_z)$ be the corresponding velocity field. Let  $\textbf{u}=\textbf{v}-\textbf{v}^{B}$ be the velocity perturbation to the Blasius solution $\textbf{v}^{B}$, itself a self-similar two-dimensional solution of the  incompressible boundary layer equations for the rescaled variable $\eta=y(U_{\infty}x)/\nu$. A local Reynolds number can be defined as $Re_{\delta^*}=U_{\infty}\delta^*(x)/\nu$, where $\delta^*=\int_0^{\infty}(1-v_x^B)/U_{\infty}\mathrm{d}y$ is the thickness of the undisturbed Blasius flow.

Direct numerical simulation of the incompressible Navier--Stokes equations is performed using SIMSON \citep{chevalier2007simson}. The equations are solved in the velocity--vorticity formulation using a pseudo-spectral method, and are advanced in time using an explicit low storage fourth-order Runge--Kutta method for the non-linear term and a second-order implicit Crank--Nicolson method for the linear terms. The velocity field $\textbf{v}(x,y,z,t)$ is expanded in $N_x$ and $N_z$ Fourier modes in the $x$ and $z$ directions, respectively. The wall-normal expansion is based on $N_y$ Chebyshev modes. To satisfy the periodicity required by the Fourier expansion in the streamwise direction despite the spatial development, a fringe region is considered at the downstream end of the computational domain. Within the fringe an artificial volume force damps all velocity disturbances and re-establishes the correct inflow profile (see appendix \ref{appA} for details). The boundary layer thickness at the inlet of the domain $\delta^*_0$ is chosen so that $Re_{\delta_0^*}=300$. Velocities and lengths are made non-dimensional by using respectively the quantities $U_{\infty}$ and $\delta^*_0$. In these units the computational domain is a rectangular box $V$ of size $(L_x,L_y,L_z)=(6000,60,100)$. The spectral resolution is $(N_x,N_y,N_z)$=$(6144,201,256)$, excluding the additional modes in $x$ and $z$ used for dealiasing by the 3/2-rule. A no-slip/no-penetration boundary condition is enforced at the wall ($y=0$) and a Neumann condition is used in the free-stream ($y=L_y$) for the three velocity components. The local resolution is comparable to the one used in \cite{duguet2012self}. 

\subsection{The moving box technique}

Numerical simulation of localized coherent structures usually require computational domains at least one order of magnitude larger than their typical size \citep{duguet2009localized}. When such coherent structures are allowed to move within the domain, length requirements become even more severe, while most of the flow domain contains only laminar, largely undisturbed flow. 

\begin{figure}
  \centering
  \includegraphics[width=1\textwidth]{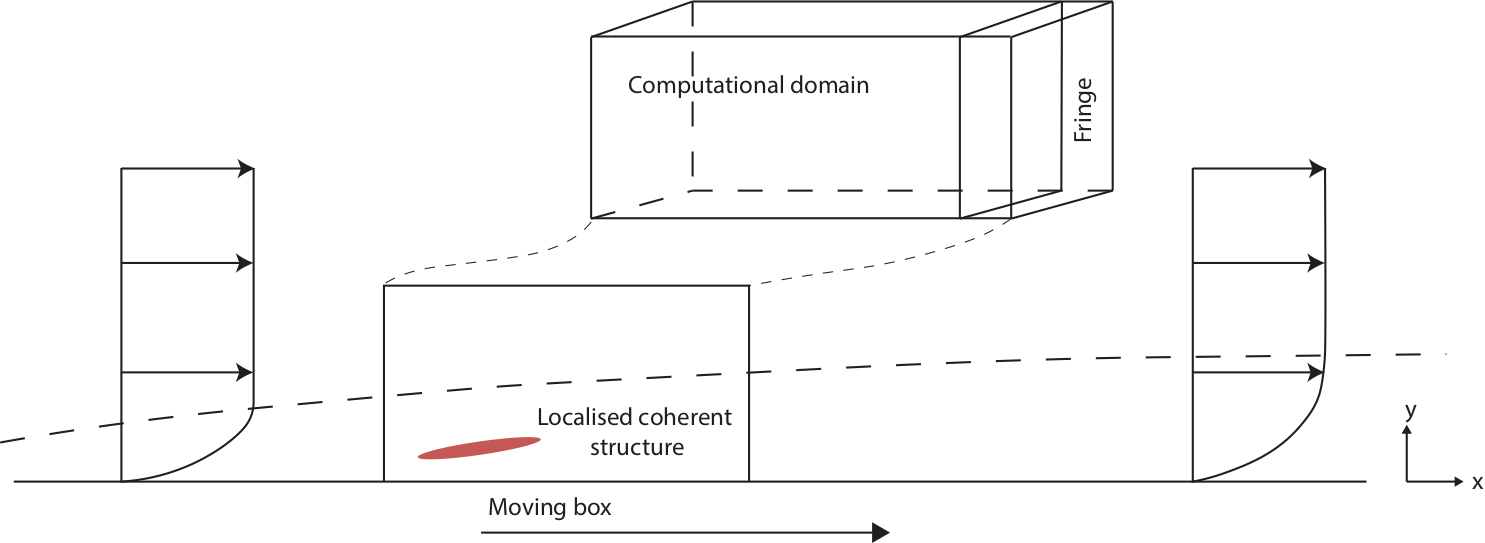}% Images in 100% size
  \caption{Sketch of the flow geometry including the moving box.}\label{fig:movingBox}
\end{figure}

In order to circumvent this issue, we suggest a moving box technique, not used previously in the case of spatially developing flows, in which the computational domain moves in the streamwise direction with piecewise constant velocity $c_{box}$ (see figure \ref{fig:movingBox}). The method then consists of Galilean changes of the reference frame of the type $v_x \leftarrow v_{x}-c_{\text{box}}$, where $c_{box}$ is the instantaneous speed of the moving box (see appendix \ref{appA} for details). In the present simulations, the box velocity takes alternatively the values $c_{box}=0$ or $0.8$ depending on the time elapsed. Different histories of $c_{box}$ yield exactly the same results as long as all travelling coherent structures remain within the computational domain. The present computation of the edge trajectory using the moving box technique was validated against the case of a non-moving domain of size $(L_x,L_y,L_z)=(12000,60,100)$, simulated with a spectral resolution of $(N_x,N_y,N_z)$=$(12288,201,256)$. 

\subsection{Edge tracking algorithm}

The dynamics on the edge is tracked iteratively using the bisection technique originally introduced by \cite{itano2001dynamics} and \cite{skufca2006edge}. A single scalar time-dependent observable $a(t)$ is selected, together with two threshold values $a_L$ and $a_T$ allowing to discard unambiguously  trajectories evolving towards the laminar or the turbulent regime, respectively. The choice of the observable is here 
\begin{equation}
%a=\left(\int_{V}|\omega_x|^2 \mathrm{d}v\right)^{\frac{1}{2}}
a=\left(\frac{1}{vol(V)}\int_{V}|\omega_x|^2 \mathrm dv\right)^{\frac{1}{2}}
\label{defa},
\end{equation}
a normalised integral measure of the streamwise vorticity $\omega_x$ in the computational domain of volume $vol(V)$, where $\mathbf{\omega}=(\omega_x,\omega_y,\omega_z)$ is the vorticity field. This observable is not affected by the choice of the reference frame. It is a measure of the amplitude of streamwise vortices, known to play a crucial role in the streak formation. It is zero for the laminar Blasius profile. Moreover the observable is also zero when the aforementioned profile is perturbed by TS waves in their linear stage (with no spanwise dependence and zero spanwise velocity). For a trajectory starting from a given initial perturbation velocity field $\textbf{u}_{0}$, localised in space, the observable $a(t)$ is monitored until its value crosses one of the two thresholds, $a_L$ from above, or $a_T$ from below. In the present study we used und by bisection, the two dotted lines are the observable bounds $a=a_L=8.74\times 10^{-5}$ and $a_T=2.68\times 10^{-3}$, though different values of $a_L$ have been tested without significant influence Depending on which threshold is crossed, the field $\textbf{u}_{0}$ is re-labelled respectively $\textbf{u}_L$ and $\textbf{u}_T$. From the history of the bisection and the knowledge of the last fields $\textbf{u}_L$ and $\textbf{u}_T$, a new update $\textbf{u}_{0} \leftarrow \frac{1}{2}(\textbf{u}_L+\textbf{u}_T)$ is constructed (for simplicity one can start with $\textbf{u}_L={\bm 0}$). This is a convergent iterative process that brackets the edge manifold.
The iterative bisection process yields a sequence of positive numbers $\lambda^{(n)}$, $n=1,2,...$ and an associated sequence of initial conditions $\textbf{u}^{(n)}$. For $n$ large enough $\lambda^{(n)}\approx \lambda^{*}$, such that in the $n \rightarrow \infty$ limit $\textbf{u}_{\lambda^{*}}=\lambda^{*}\textbf{u}^{(0)}_{0}$ lies exactly on the edge manifold $\Sigma$, and $\textbf{u}_{\lambda^{(n)}}$ is an approximation of such a state. In practice, when machine precision is met regarding the accuracy of $\lambda^*$, the whole bisection process is restarted from another state further along the edge trajectory. This bisection technique is similar to that used in previous subcritical systems, except that crossing the bound $a_L$ is here \emph{not} necessarily interpreted as relaminarisation. In \cite{cherubini2011edge} and \cite{duguet2012self} the time horizon in the bisection was not sufficiently long for the linear instability of the base flow to manifest itself, therefore crossing the threshold $a_L$ from above necessarily corresponded to an effective relaminarization. We demonstrate here that this picture needs be revised when bisecting is no longer limited in time. 

\section{Results}

\subsection{The edge for moderate times}

\begin{figure}
\centering
  \includegraphics[width=4.5cm]{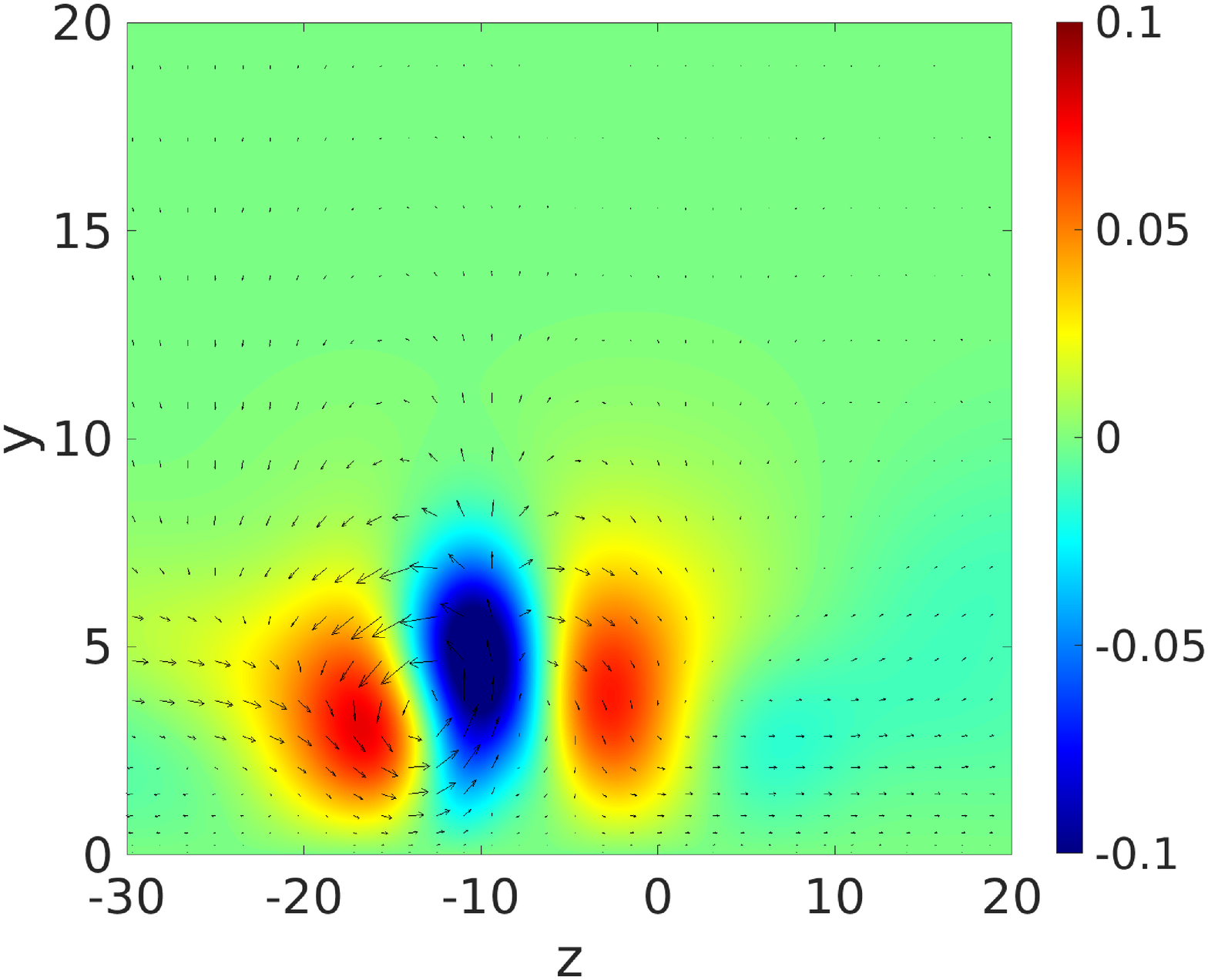}% Images in 100% size
  \includegraphics[width=4.5cm]{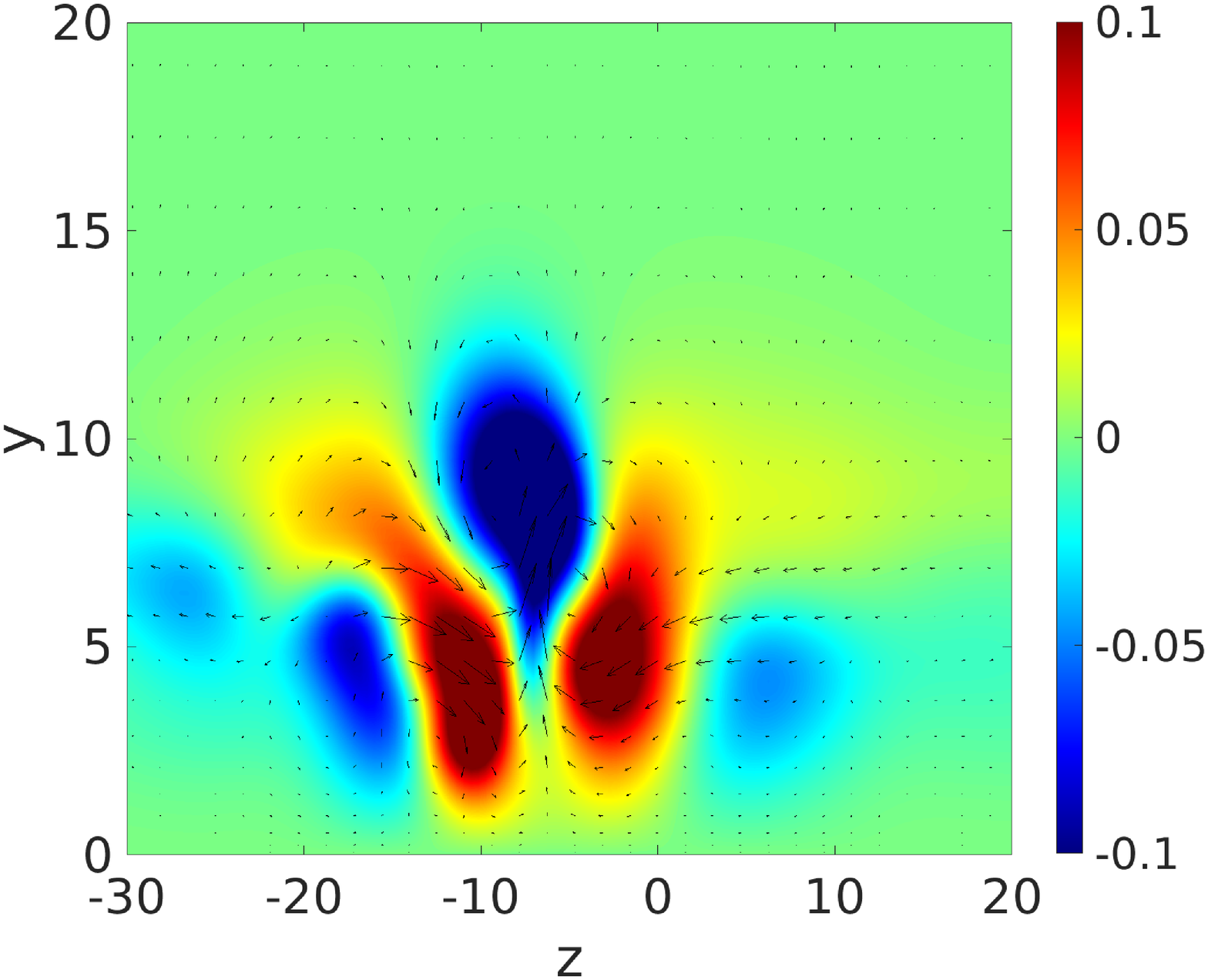}% Images in 100% size
  \includegraphics[width=4.5cm]{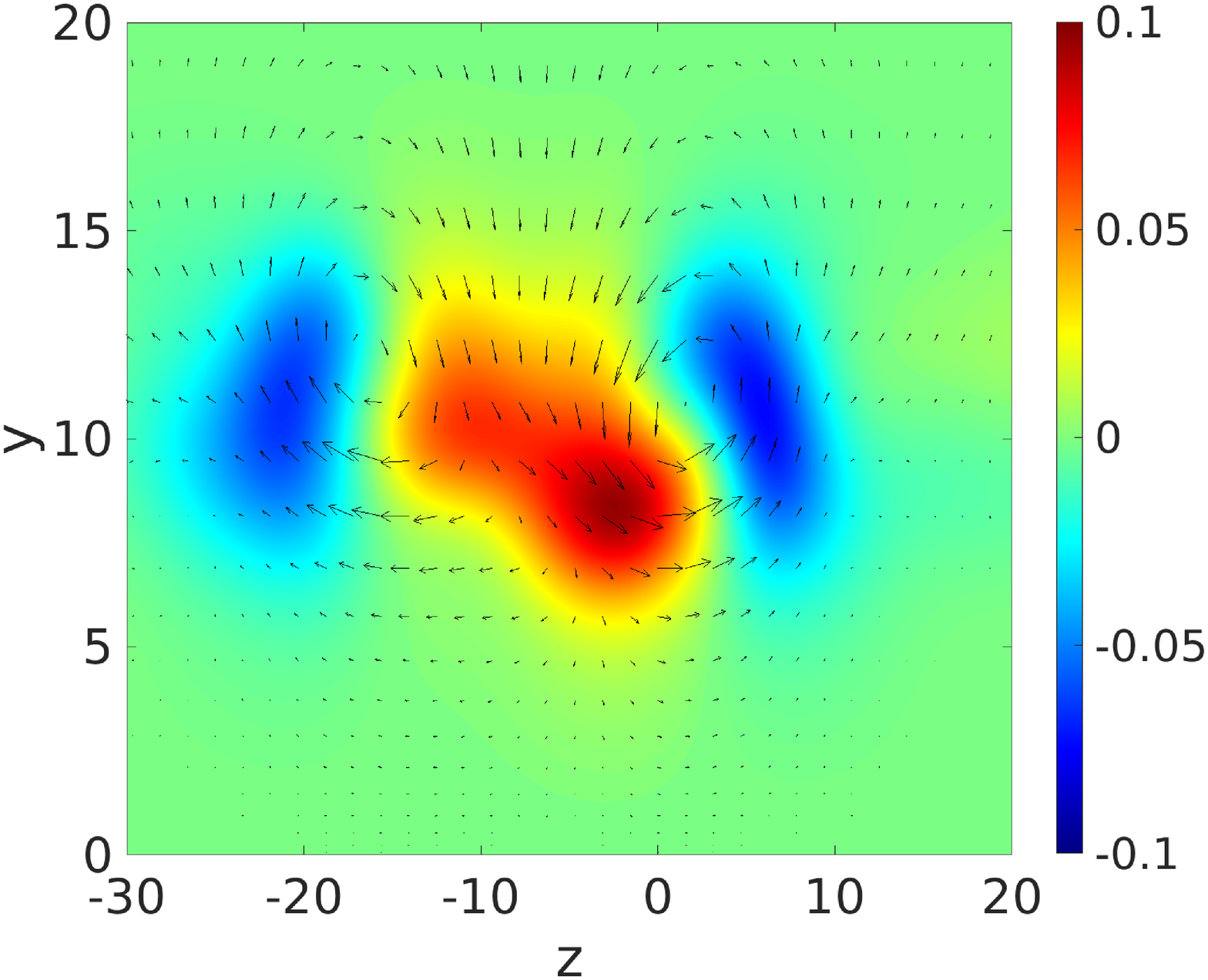}% Images in 100% size
  \caption{Illustration of a streak switching event along the edge trajectory at $t=3550$. $(y,z)$ cross-sections of $u_x$. From left to right $x=2140,\ 2420$ and $2980$.} \label{fig:cross_section}
\end{figure}

The initial condition $\textbf{u}_0$ consists of a pair of counter-rotating vortices located close to the wall at $x = 50$, $z=0$, and $\emph{not}$ aligned with the main flow direction; unlike in \cite{duguet2012self} the evolution of the flow is here free from any discrete symmetry constraint.
\begin{figure}
\centering
  \includegraphics[width=14cm]{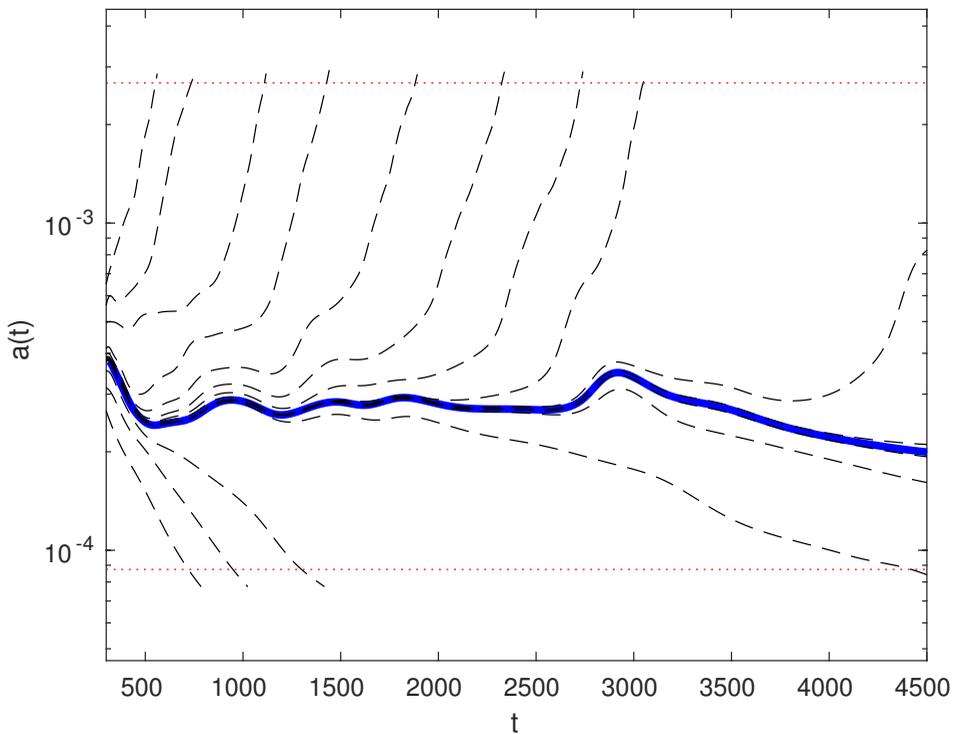}% Images in 100% size
  \caption{Observable $a(t)$ vs. $t$ during application of the bisection algorithm over moderate time horizons. The thick line (blue online) represents the edge trajectory found by bisection, the two dotted lines are the observable bounds $a=a_L=8.74\times 10^{-5}$ and $a_T=2.68\times 10^{-3}$ (see text).}
  \label{fig:atshort}
\end{figure}
For  $0\le t \le 4700$ (in units of $\delta_0^*/U_{\infty}$) it is possible to effectively bracket the edge using the algorithm described above and the current definition of the observable $a(t)$. Time series of $a(t)$ during the bisection process are shown in Fig. \ref{fig:atshort}.
\begin{figure}
\centering
  \includegraphics[width=0.95\textwidth]{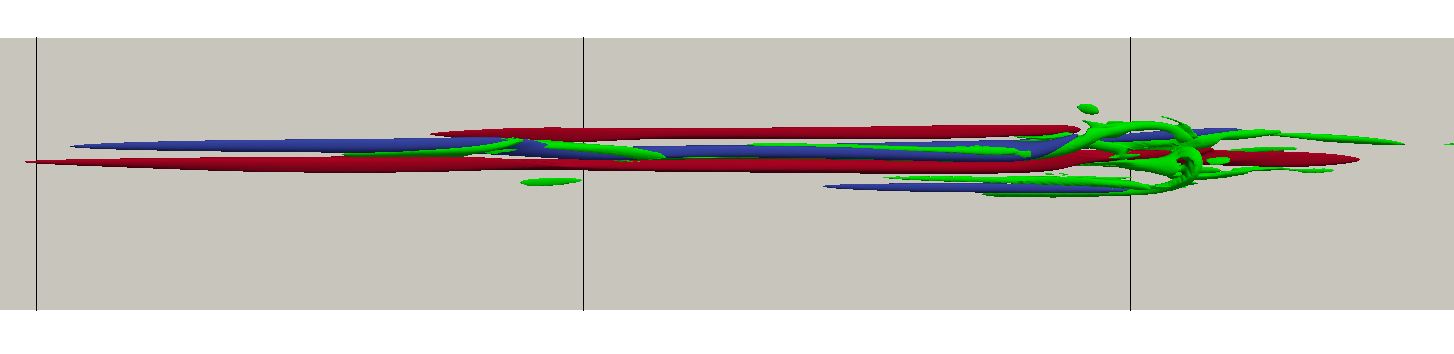}
  \includegraphics[width=0.95\textwidth]{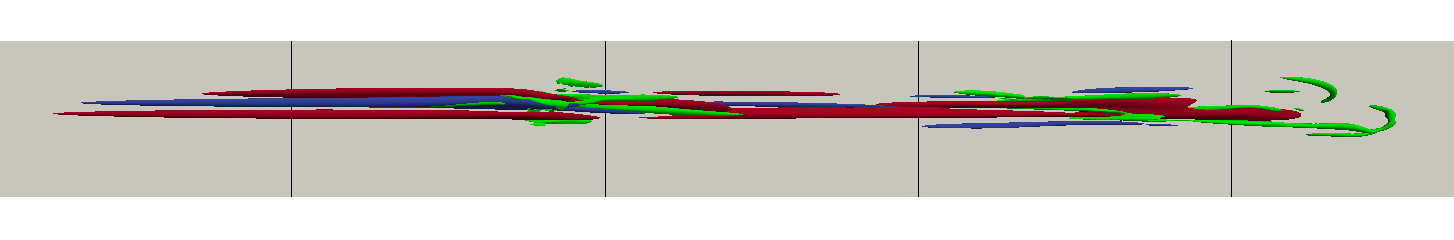}
  \includegraphics[width=0.95\textwidth]{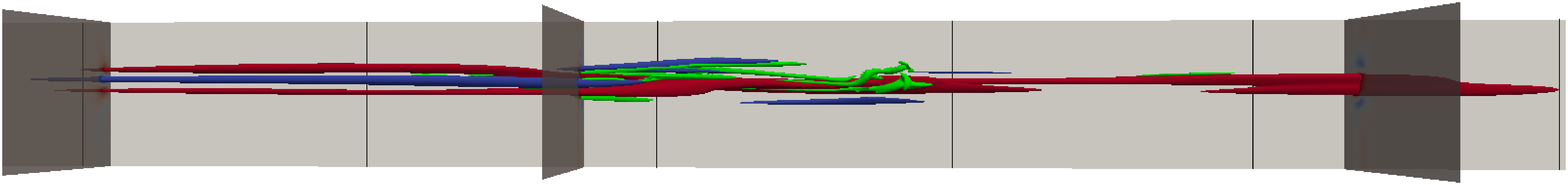}
  \caption{Three-dimensional perspective view from above of the edge trajectory at times $t=1700,\ 2750$ and $3550$ (from top to bottom), showing isosurfaces of streamwise perturbation velocity with respect to spanwise mean with values $0.06$ and $-0.08$ (red and blue, respectively), together with vortical structures $\lambda_2=-1.5\cdot 10^{-5}$ (green). Flow from left to right. The black lines are separated by a distance of 200 in units of $\delta_0^*$. The last snapshot includes the cross sections presented in figure \ref{fig:cross_section}} \label{fig:edgeSnaps}
\end{figure}
The dynamics along the edge trajectory is globally comparable to that in \cite{khapko2016edge} though there are also clear differences. The velocity field along the edge trajectory is characterized by long streamwise streaks and streamwise vortices, both characteristics of a self-sustained process described in \cite{duguet2012self}. It remains spatially localized in three dimensions at all times. Three-dimensional visualisations of three specific snapshots are shown in figure \ref{fig:edgeSnaps}, where a typical regeneration cycle of the edge dynamics is displayed. The whole localized structure is unsteady;  similarly to \cite{duguet2012self} it is possible to identify a robust streaky core upstream, together with recurrent secondary structures that detach from the main core, travel further downstream and dissipate. The global structure and the temporal dynamics along this edge trajectory is qualitatively closer to that computed in other flows, notably the localised edge state obtained in the ASBL \citep{khapko2016edge} except for the sparser occurrence of bursts. Compared to the spanwise-symmetric analyzed in \cite{duguet2012self}, the streak organisation is different. The robust pair of symmetric low-speed streaks flanking a high-speed streak is lost in favour of an unsteady dynamics, where the sandwiched streak is alternatively the high-speed and the low-speed one. As this process does not occur simultaneously at all locations, it is possible to visualise different stages of the streak switching cycle at the same time by comparing for instance the initial and final frames in figure \ref{fig:cross_section}. Visualisation of the $\lambda_2$ criterion \citep{Jeong95} in figure \ref{fig:edgeSnaps} shows the presence of slightly asymmetric vortical structures above the regions where the streaks pinch together. Our observations suggest that the strongest of these approximately hairpin-shaped vortices, because of the wall-normal flows they induce,  act as precursors of the switch events such as the one in figure \ref{fig:cross_section}. Streak switching appears recurrently in the literature as characteristic of unsteady edge states (\cite{tohitano2003,khapko2013localized,khapko2016edge,biau2012laminar}), and could now be confirmed in spatially developing boundary-layer flows as well.

%\rev{Three snapshots of the cross-section of the edge trajectory describing the switching phenomena at $t=3550$ can be found in figure \ref{fig:cross_section}. Scanning the $u'$ in the coherent structure downstream shows that it locates further away from the flat plate as we move in the $x$ direction.}

The spatially developing nature of the boundary layer suggests the use of a 'local' boundary layer thickness as a characteristic length scale. To define a lengthscale rigorously it is necessary to refer to an unambiguous streamwise location. Exploiting the spatial localisation of the coherent structure under study, we define its streamwise position $x_G$ in analogy with the centre of mass by
\begin{equation}
    x_{G}(t) = \frac{\iint  x\ |u_{x}(x,y_p,z,t)|\ \mathrm{d}x\mathrm{d}z}{\iint  |u_{x}(x,y_p,z,t)|\ \mathrm{d}x\mathrm{d}z}, \label{eq:xG}
\end{equation}
where $y_p$ is the wall-normal coordinate of a given plane known to intersect the coherent structure at all times (in practice $y_p=1.5\delta_0^{*}$). The weighting function in the integrals in equation \eqref{eq:xG} is the absolute value of the streamwise perturbation velocity $u_x$,  $x_G(t)$ is an estimation of the displacement of the main coherent structure with time from its initial location. The local displacement thickness $\delta=\delta^*(x_G)$, evaluated at this time-dependent location, provides the scaling factor $\delta/\delta^*_0$ for the boundary layer growth. 

\begin{figure}
  \centering
  \includegraphics[width=10cm]{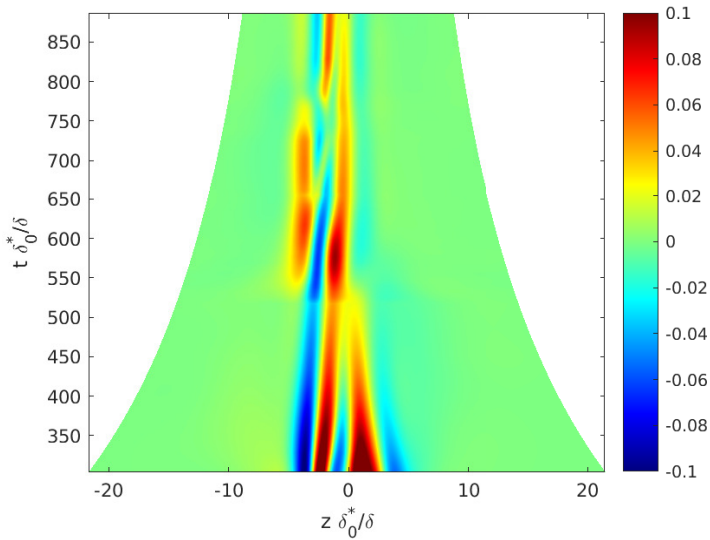}% Images in 100% size
  \caption{$(z,t)$ space-time diagram of $u_x(x,y,z,t)$ along the edge trajectory, in a frame moving with the center of mass located at $x=x_G(t)$ and $y=y_p$, of the perturbation streamwise velocity. Both $z$ and $t$ are rescaled by the local boundary layer thickness $\delta/\delta^*_0$, causing the apparent narrowing of the domain.}\label{fig:spaceTime_edge}
\end{figure}

Monitoring $x_G(t)$ and its temporal derivative $\dot{x}_G(t)$ is a convenient way to define the advection speed of the structure as a whole. After a relatively short initial transient of $t\approx 500$, $\dot{x}_G(t)$ reaches an almost constant value of 0.625 in units of $U_{\infty}$, comparable to the values found in \cite{duguet2012self}. Velocities scale with $U_{\infty}$ and are unaffected by the rescaling, while vorticities and the domain dimensions are now time-dependent. The rescaled space--time diagram for the streamwise velocity perturbation $u_x(x=x_G(t),y=y_p,z,t)$ (evaluated at the center of mass of the coherent structure) is shown in figure \ref{fig:spaceTime_edge}. The approximately constant width of the streak core in time is a confirmation that the $\delta^*(x_G)/\delta^*_0$-scaling is relevant at least for the spatial coordinates. Similarly, the same scaling for the time coordinate suggests that the apparent period for the streak switching becomes inherently larger as the boundary layer grows. This diverging timescale, together with the diverging length of the streaks, is the main quantitative obstacle to significantly longer edge tracking. In practice it is possible to observe only two full streak switching cycles, during the time intervals $t\approx$ 600--1300 and 1900--3300, \textit{i.e.}\ in the rescaled variables $t \delta_0^*/\delta\approx$ 300--420 and 630--770. Furthermore the whole state grows in length, from $\approx 400\delta^*_0$ at $t=1750$ to $\approx 1200\delta^*_0$ at $t=4700$, as shown in figure \ref{fig:edgeSnaps}.

\subsection{Coexistence of Tollmien--Schlichting waves and streaks for long times}

\subsubsection{Physical space}

The edge trajectory is by construction linearly unstable. So is the laminar base flow, except that the instabilities manifest themselves over different timescales, with the base flow instability being typically slower \citep{zammert2017transition}. If the time horizon in the bisection is large enough there is a competition between bypass transition (expected from the linear instability of the edge) and the classical transition scenario (featuring Tollmien--Schlichting waves, due to the linear instability of the base flow). We report now the unforced simultaneous occurrence of these two transition mechanisms.
%\rev{We define here an early leaver as the trajectory which leaves too the edge early enough so that both transition mechanisms have not manifested together}.

\begin{figure}
\centering
  \includegraphics[width=7cm]{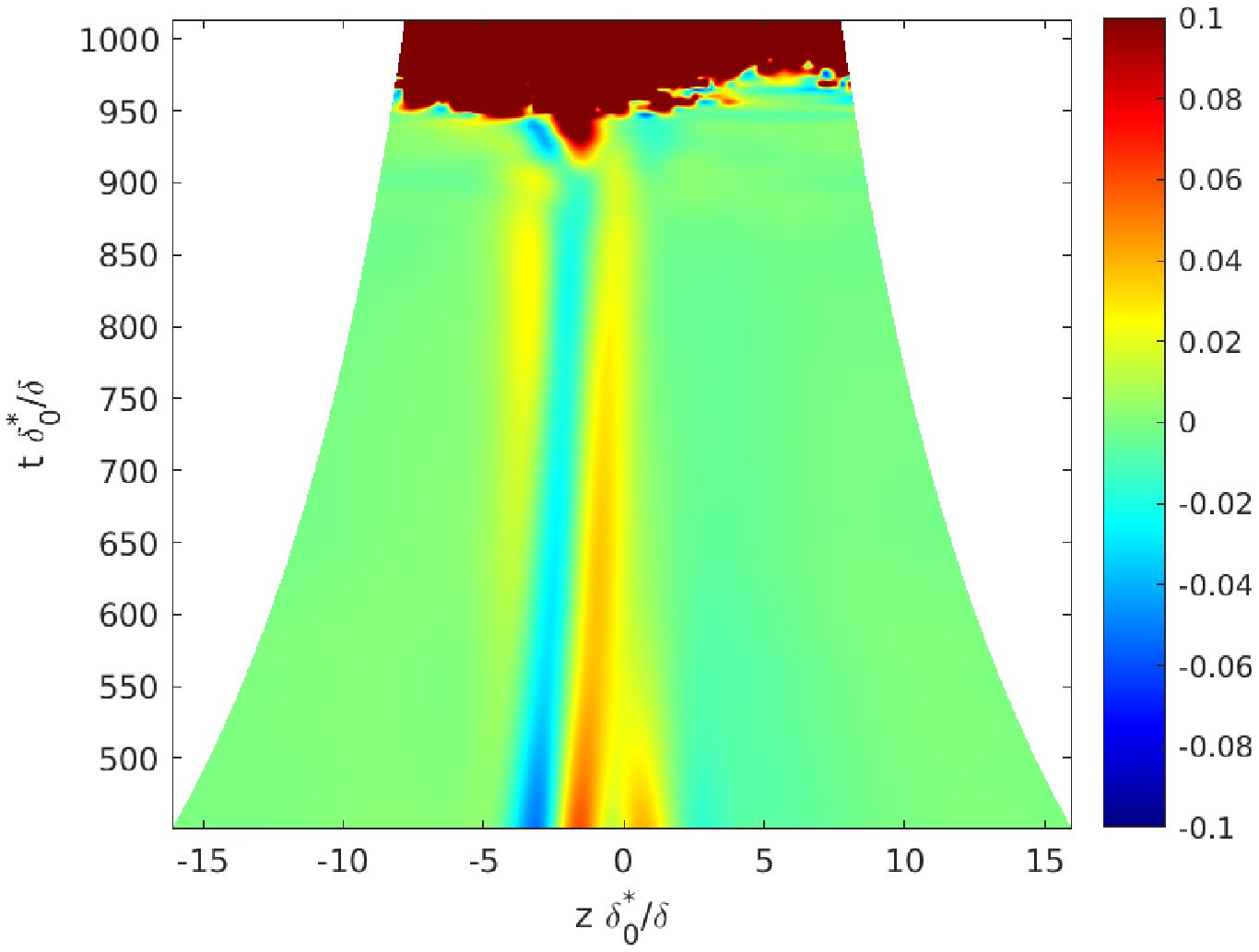}% Images in 100% size
  \includegraphics[width=7cm]{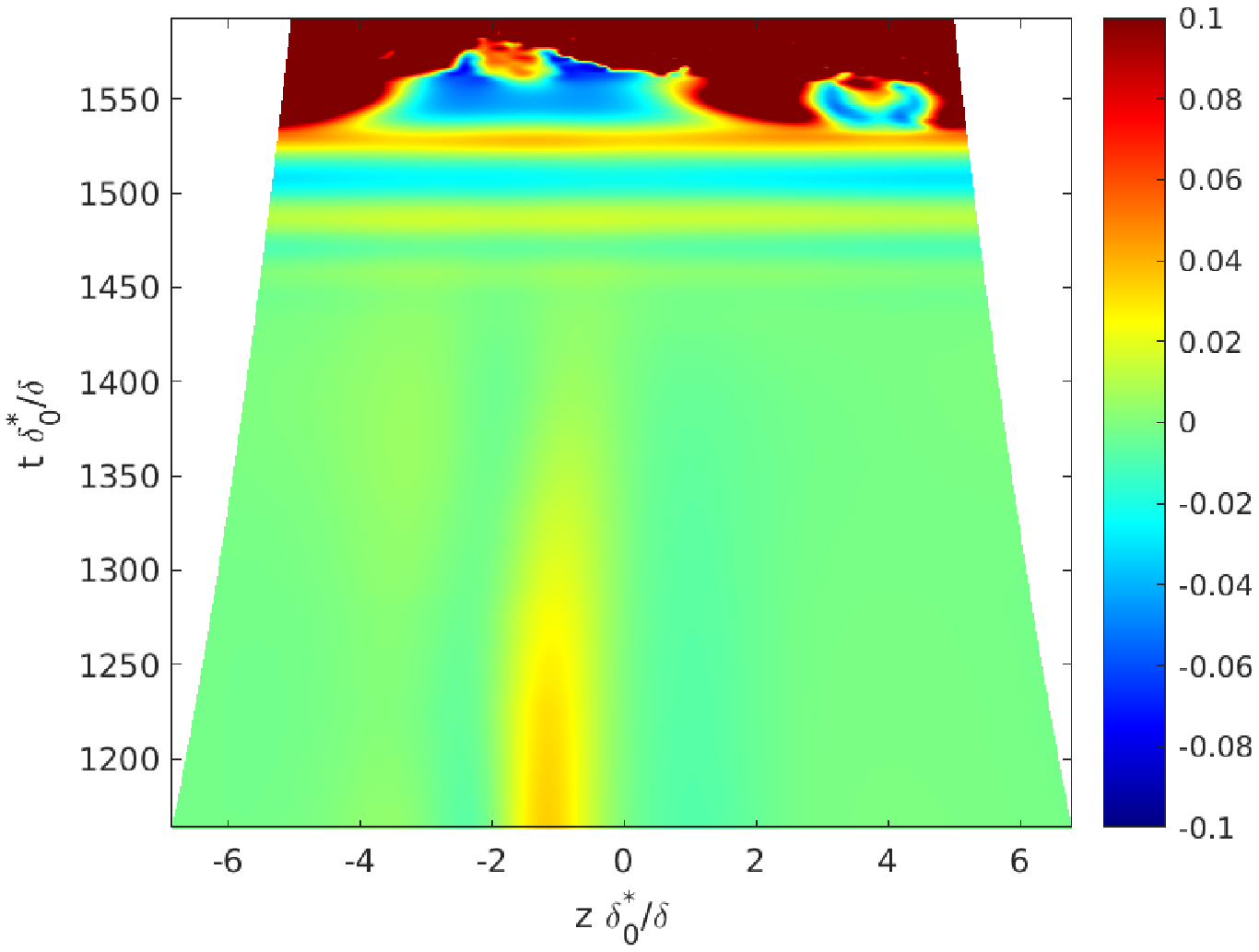}% Images in 100% size
  \caption{$(z,t)$ space--time diagrams of $u_x(x,y,z,t)$ evaluated at $x=x_G(t)$ and $y=y_p$, for two initially nearby trajectories bracketing the edge trajectory. Both $z$ and $t$ are rescaled by the local boundary layer thickness $\delta/\delta^*_0$. Left:  bypass transition route. Right: streak decay followed by classical transition based on the growth of the Tollmien--Schlichting waves.} \label{fig:transitionScenarios}
\end{figure}

\begin{figure}
\centering
  \includegraphics[width=1.1\textwidth]{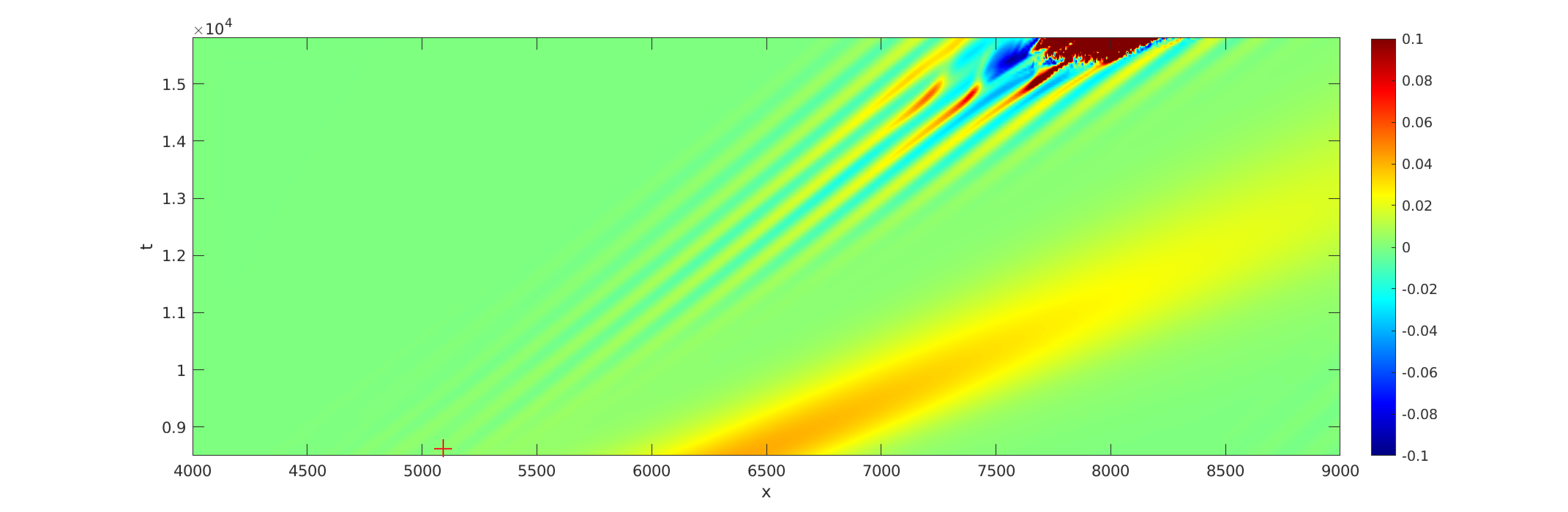}
  \caption{Space-time diagram of $u_x(x,y,z,t)$ for $z=-10$ and $y=y_p$ for the trajectory away from the edge manifold, as displayed in figure \ref{fig:transitionScenarios} (right). The streaks decay while the TS wavepacket grows in amplitude, forming a turbulent spot. The red cross indicates the initial position of the crest tracked in figure \ref{fig:speeds}.} \label{fig:TSwaves}
\end{figure}

\begin{figure}
\centering
  \includegraphics[width=6.7cm]{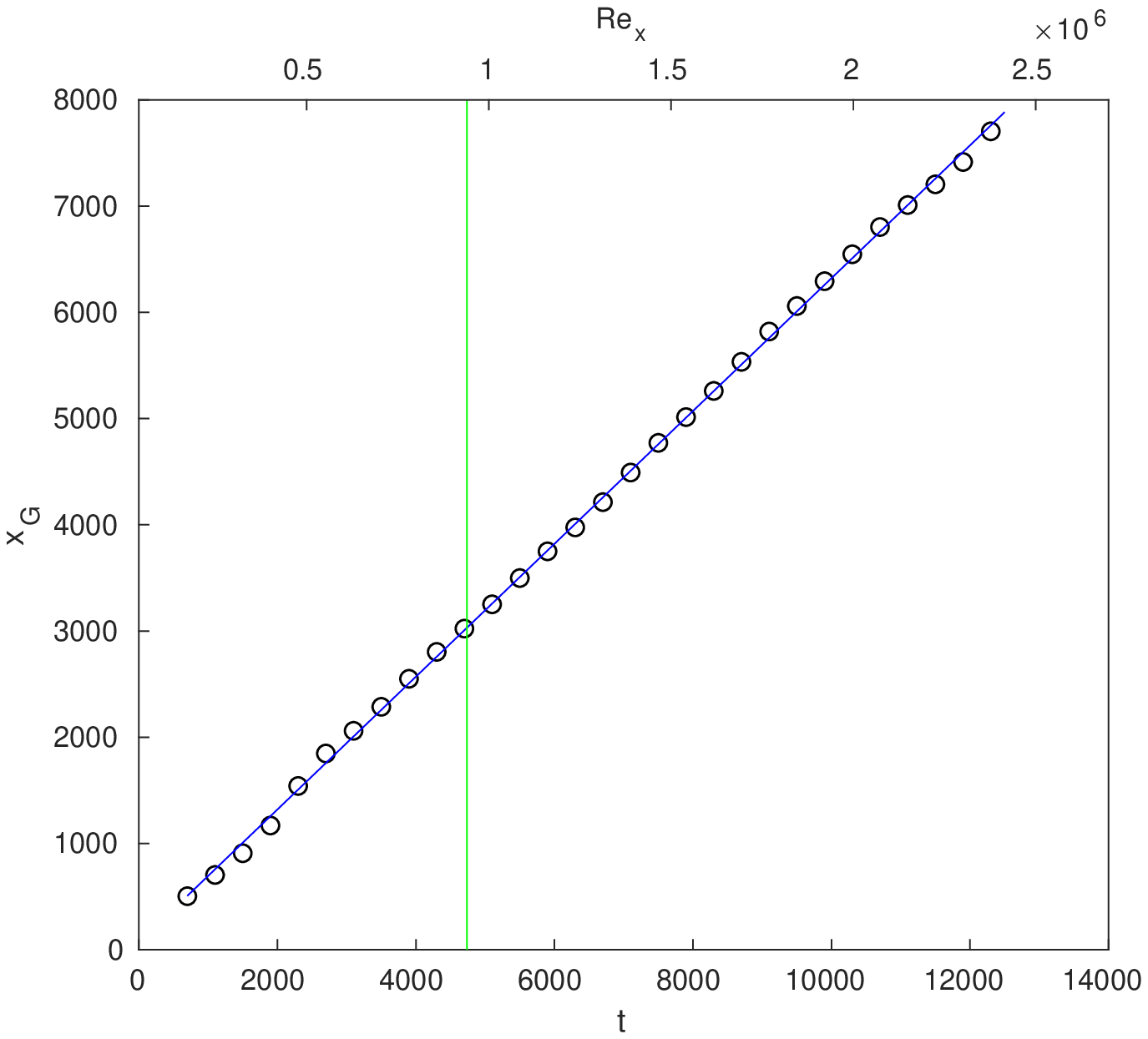}
  \includegraphics[width=6.6cm]{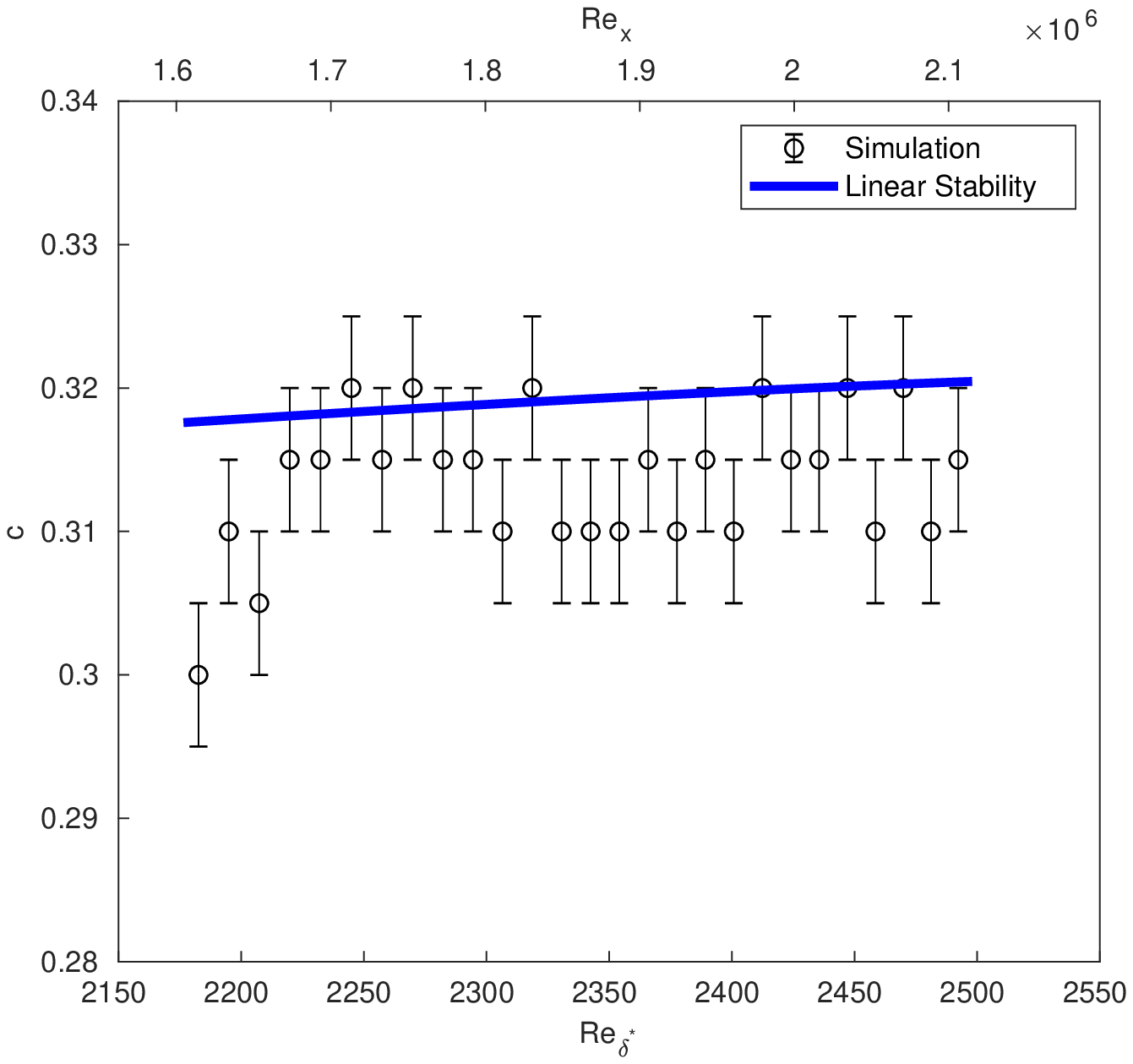}
  \caption{Streamwise propagation velocities. Left: position of the center of mass of the coherent structure $x_G(t)$ for a trajectory below the edge versus time. The vertical line indicates the time limitation of edge tracking (see text). Circles: data, solid line (blue online): linear fit. Right: phase velocity of TS waves, numerical simulation vs. linear instability analysis (see text). The error bars represent the accuracy in estimating the convection speed from figure \ref{fig:TSwaves}.} \label{fig:speeds}
\end{figure}

 Figure \ref{fig:transitionScenarios} shows that a relative difference $\delta \lambda/\lambda^{*}$ of magnitude $\mathcal{O}(10^{-4})$ in the choice of the bisection parameter $\lambda=\lambda^* + \delta \lambda$ leads to different transition scenarios in the long-time horizon. For $\delta \lambda>0$ destabilization of the edge trajectory leads to breakdown of the streaks akin to that reported in bypass transition \citep{brandt2004transition}. The breakdown of the streak is local in $z$, as visible in figure \ref{fig:transitionScenarios} (left). On the other hand, for $\delta \lambda<0$ the streaky structure is not strong enough to self-sustain, instead it starts to decay viscously. During the streak decay the linear instability of the flow manifests itself in the form of TS waves emerging upstream of the streak core, well visible in figure \ref{fig:transitionScenarios} (right). The apparent wavelength as well as the propagation speed of these waves makes it clear that they correspond to TS waves. Regarding wavelengths, these waves have $k_z=0$ and $k_x<0.35$ at all times, which matches the linear stability curve from the literature \citep[\textit{cf.}][]{BerlinPhD}. The speed of propagation of the waves can be measured from space--time diagrams such as figure \ref{fig:TSwaves}. It is approximatively 0.32 $\pm$ 0.005, matching quantitatively the speed of the TS waves obtained from linear stability analysis of the frozen Blasius profile at their onset \citep{schmid2012stability}. A more detailed study of the TS waves is shown in figure \ref{fig:speeds} (right). The figure shows the comparison between the nonlinear TS waves in our simulation and the results from linear stability analysis, for a streamwise wavenumber $k_x\approx 2\pi/\lambda_{TS}=0.037$, where $\lambda_{TS}=170\delta_0^*/\delta_{x,TS}^*$. The wavelength $\lambda_{TS}$ is obtained from the intercrest distance 170 scaled with the local boundary thickness $\delta_{x,TS}^*$, itself an evaluation of $\delta^*$ at the position $x=x_{TS}$. The position $x_{TS}$ is obtained by tracking in time and space the specific crest passing through $x=5186$ for time $t=8500$ (indicated in figure \ref{fig:TSwaves} by a cross). Moreover figure \ref{fig:speeds} (left) shows the position of $x_G(t)$ from the trajectory bracketing the edge from below. The position of the edge is well fitted by the straight line $x_G(t)=x_{G0}+v_{Gx}t$, where $x_{G0}=70$ and $v_{Gx}=0.625$. It is clear that the TS waves travel at a velocity approximately half that of the streak core. This slower propagation speed makes the associated TS wavepacket detach progressively from the decaying streak core, while as a result the position of the center of mass drifts upstream of the streak core as seen in figure \ref{fig:transitionScenarios} (right). The exact way how the TS wavepacket is triggered is not fully understood, preliminary examination points towards disturbances growing from the wake of the streaks. The waves grow upstream of the coherent structure, and they do not interact with the streak core. This is consistent with previous observations that finite-amplitude streaks act as additional damping of TS-wave growth \citep{cossu2004tollmien} and that streaks and TS waves are not seen to overlap spatially. The TS wavepacket grows in size as it travel downstream, with the number of individual rollers growing too \citep{gaster1975}. At later times the waves inside the wavepacket undergo a secondary instability with spanwise wavenumber $k_z \ne 0$ (of fundamental Klebanoff type), followed by a rapid local breakdown into turbulence.

\subsubsection{State space}

Further complementary information about the global temporal dynamics can be obtained from phase portraits. We use here the $L^2$-norms of the three integral observables $\omega_x$, $\omega_y$ and $u_z$. In order to take into account the evolution of the boundary layer thickness downstream and the variable size of the domain when expressed in units of $\delta(t)$ rather than $\delta_0^*$, the r.m.s. vorticities need to be rescaled using the correcting factor $(\delta_0^*/\delta)^{\frac{1}{2}}$, and the r.m.s. velocities by $(\delta_0^*/\delta)^{(3/2)}$, as explained in \cite{duguet2012self}. We hence consider the three phase portraits observables $\Omega_x$, $\Omega_y$ and $W$, defined by equation \eqref{eq:observables1}--\eqref{eq:observables3} and parametrised by time only:
\begin{eqnarray}
%\Omega_x=(\delta_0^*/\delta)^{\frac{1}{2}}\left(\int_V|\omega_x|^2dv\right)^{\frac{1}{2}},\\
\Omega_x=(\delta_0^*/\delta)^{\frac{1}{2}}\left(\frac{1}{vol(V)}\int_V|\omega_x|^2\mathrm dv\right)^{\frac{1}{2}},
\label{eq:observables1}\\
%\Omega_y=(\delta_0^*/\delta)^{\frac{1}{2}}\left(\int_V|\omega_y|^2dv\right)^{\frac{1}{2}},\\
\Omega_y=(\delta_0^*/\delta)^{\frac{1}{2}}\left(\frac{1}{vol(V)}\int_V|\omega_y|^2\mathrm  dv\right)^{\frac{1}{2}},
\label{eq:observables2}\\
%W=\left(\int_V|u_z|^2dv\right)^{\frac{1}{2}},
W=(\delta_0^*/\delta)^{\frac{3}{2}}\left(\frac{1}{vol(V)}\int_V|u_z|^2\mathrm dv\right)^{\frac{1}{2}},
\label{eq:observables3}
\end{eqnarray}

\begin{figure}
\centering
  \includegraphics[width=12cm]{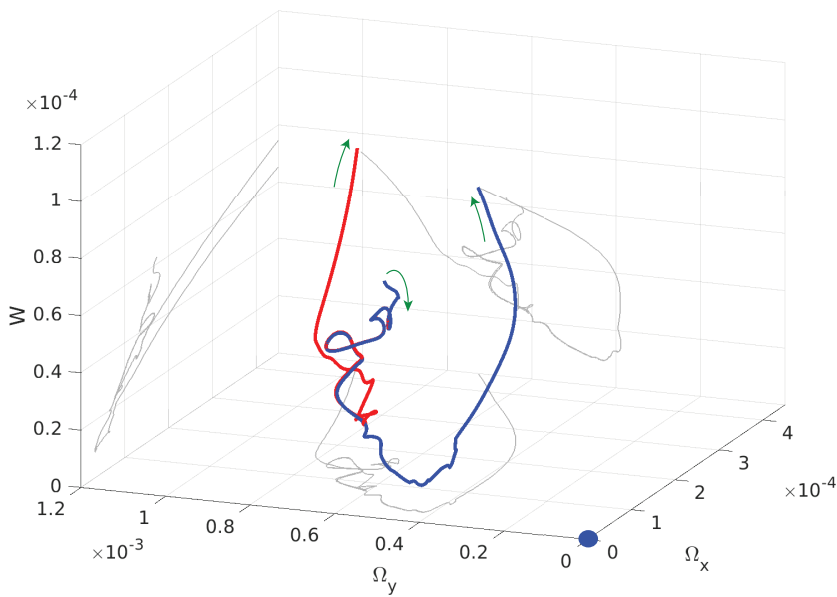}% Images in 100%
  \caption{Three-dimensional phase portrait using global variables $\Omega_x$, $\Omega_y$ and $W$. The initial condition associated lies at approximately ($\Omega_x$, $\Omega_y$, $W$) = (1.72, 6.62, 1.80)$\times 10^{-4}$ but the first 500 time units are not shown. The blue dot at the $(0,0,0)$ is the laminar state. The red trajectory corresponds to bypass transition, the blue one to classical TS transition, and their points in common define the converged part of the edge trajectory. In grey the projections on the different two-dimensional planes.} 
  \label{fig:phase_space}
\end{figure}

\begin{figure}
\centering
  \includegraphics[width=12cm]{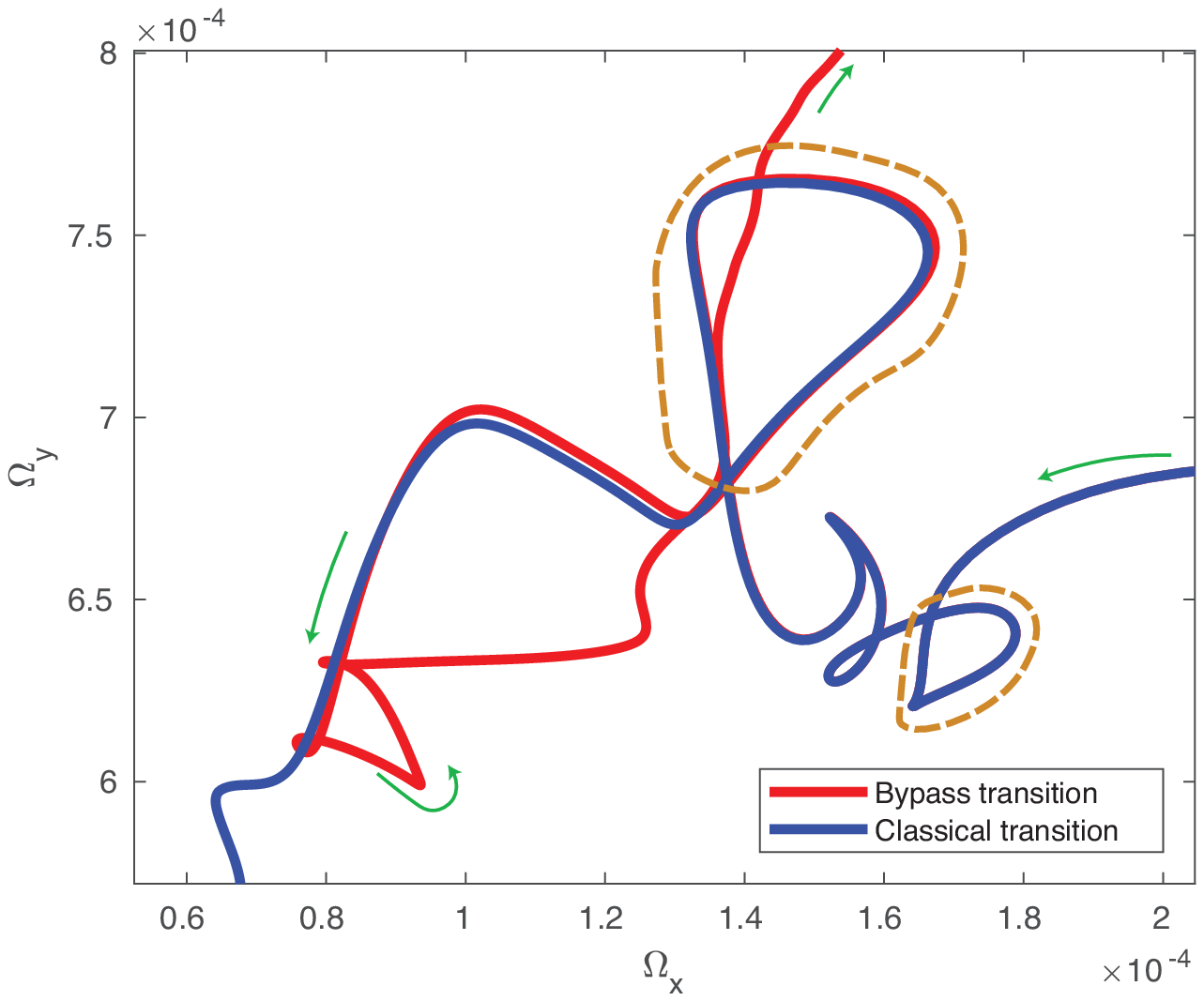}% Images in 100% size
  \caption{Zoom on the edge region using only $\Omega_x$ and $\Omega_y$ (same data as Figure \ref{fig:phase_space}). The dotted lines correspond to two other bracketing trajectories. The first 300 time units are omitted. Swirls are indicated using closed dashed lines (orange online).}
  \label{fig:phase_space2D}
\end{figure}

The reduced phase portrait is displayed in figure \ref{fig:phase_space} (left). The initial condition associated with $\lambda^{*}$ corresponds to the point ($\Omega_x$, $\Omega_y$, $W$) =(1.72, 6.62, 1.80)$\times 10^{-4}$. Trajectories bracketing the edge have an initial condition very close to it, and are indistinguishable in the figure. Past a initial transient of approximatively $500\delta^*_0/U_{\infty}$, all trajectories  approach a single recurrent region bounded from above by $W=6\times 10^{-5}$ and characterised by apparent swirls in all displayed variables. At later times the trajectories leave this part of state space and evolve towards higher $W$ values associated with turbulent flow. The recurrence region corresponds to the neighbourhood of the unstable edge state.  It is displayed in the zoom in figure \ref{fig:phase_space2D}, where up to three recurrent cycles can be distinguished. Each of the swirls in the ($\Omega_x$, $\Omega_y$) plane corresponds in physical space to a streak switching event and a release of vortical perturbations downstream. The dynamics of the trajectories leaving the edge is interesting, because independently of the route to turbulence all trajectories point towards the same direction, \textit{i.e.}\ towards a common turbulent attractor. This last aspect, sets this flow apart from the usual subcritical picture characterised by the coexistence of two different attractors with clearly distinct basins of attraction (see \textit{e.g.}\ the introductory sketch in \cite{DuguetPoF2013}). 

%The variables in the phase space are rescaled to account for the boundary layer thickness, the vorticities must incorporate a correcting factor as indicated in \cite{duguet2012self}, becoming $\Omega_i=(\delta_0^*/\delta)^{(1/2)}\omega_i^{rms}$ where $\delta = \delta^*(x_G)$\highlight{MB: I think it is better defined now considering the definitions above}. In figure \ref{fig:phase_space} we plot a three-dimensional phase-space \revdd{projection}{portrait} $(\Omega_x,\Omega_y,u_z^{rms})$ and a two-dimensional one $(\Omega_x,\Omega_y)$.

%From the phase-space it can be seen that a relative attractor is reached after an initial transient \highlight{MB: We could maybe explain this in a better way}, then there are times of higher activity, with increasing $\Omega_x,\Omega_y$ and $u_z^{rms}$. Under the appropriate projection as in figure \ref{fig:phase_space} (right) it can be seen that up to three such cycles occur, however they are not coincident due to the lack of symmetry.

%\begin{figure}
%  \centerline{\includegraphics{SpaceTime_ST_SC}}% Images in 100% size
%  \caption{Streak transition}
%\end{figure}
%\begin{figure}
%  \centerline{\includegraphics{SpaceTime_TS_SC}}% Images in 100% size
%  \caption{TS transition}
%\end{figure}

%\begin{figure}
%  \centerline{\includegraphics{phase_space_SC}}% Images in 100% size
%  \caption{Phase space}
%\end{figure}

%\begin{figure}
%  \centerline{\includegraphics{phase_spaceSC_ZOOM}}% Images in 100% size
%  \caption{Phase space}
%\end{figure}

\section{Discussion and conclusions}

Large-scale computational edge tracking in a spatially developing Blasius boundary layer has been revisited, considering much longer time horizons than previously. In addition to a better understanding of the short-time dynamics, the results call for a revision of the concept of edge at large times, in the special case where the base flow instability is of a supercritical nature rather than subcritical. On the shorter timescales, a regeneration process is confirmed, whose characteristic timescale increases as the boundary layer thickness grows. Before the edge instability, recurrent visits to a streaky active core are observed as in most previous studies, \cite{duguet2012self,khapko2013localized,khapko2016edge} including the slow phenomenon of streak switching, observed here in a spatially developing configuration. Furthermore, concomitant scenarios of bypass and classical transition have been observed on longer timescales, and their possibly simultaneous occurrence blurs the long-time output of the edge tracking algorithm. To our knowledge, whereas instances of such a coexistence have been reported in more complicated geometries \citep{xu2017destabilisation, canton2019critical} this coexistence is reported for the first time that in a simulation of an unforced boundary layer flow. 

%This coexistence of transition routes opens the door to a study of the interaction between the transition mechanism and the stable manifolds dynamics in line with \cite{zammert2017transition} and \cite{xu2017destabilisation}. 

\begin{figure}
\centering
  \includegraphics[width=12cm]{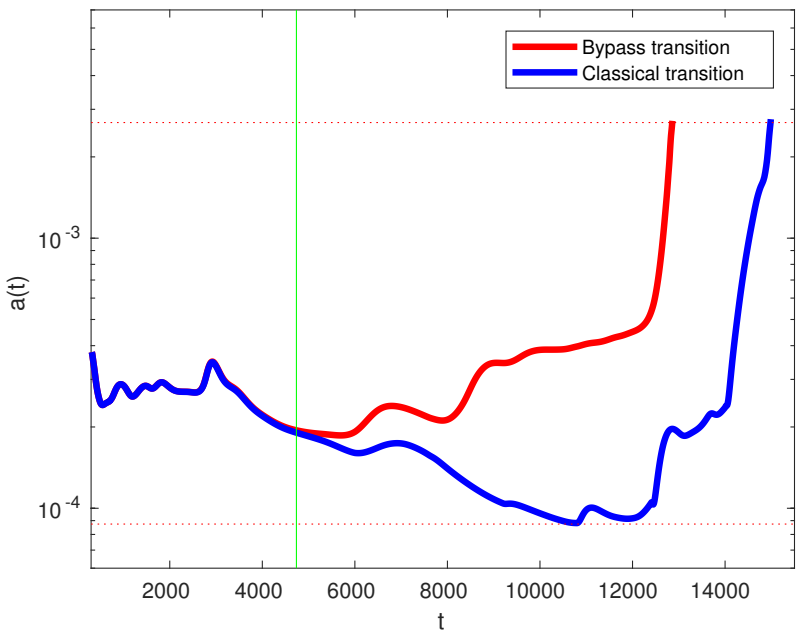}% Images in 100% size
  \caption{Observable $a(t)$ vs. $t$ during application of the bisection algorithm. The green vertical line marks the maximum time for the edge tracking algorithm using that observable, and the two dotted lines stand for the observable bounds $a=a_L$ and $a_T$ (see text).}
  \label{fig:at}
\end{figure}

A moving box technique allows for a more efficient usage of the computational domain. This moving box technique is a simple generalisation of the Galilean changes of reference frequently used in simulations of parallel flows. It proves efficient for the tracking of all spatially localized coherent structures; it is not limited to edge tracking and can also handle minimal seeds, linear or nonlinear localised wavepackets, and incipient turbulent spots. The longest individual simulations requested more than 12,000 time units for one of the two bounds $a_L$ or $a_T$ to be reached, this is at least four to six times more than in \cite{duguet2012self} due to the use of the moving box technique. If both the streaky state and the TS waves, which travel at markedly different speeds, are contained in the same computational domain (the worst-case scenario), the improvement in tracking time is 40\% compared to the fixed case, and much more otherwise. In total, edge tracking has been achieved here over a time horizon three times longer than in previous investigations of the Blasius boundary layer. The temporal limit for the edge tracking is met when the relative difference in the main observable between the closest edge-bracketing trajectories is $2\%$ or less, and we have no further information on the nature (bypass or classical) of the trajectories within these 2$\%$. This limit corresponds here $t\approx$ 4700 and the corresponding distance, expressed in non-dimensional units of $Re_x=U_{\infty}x/\nu$, is $Re_x^{max}\approx 9\times 10^5$. %Compared to \cite{duguet2012self}, it is the combination of the moving box technique with edge tracking allows here to extend the total time horizon of the bisection by a factor of $\approx$ 3.\\

From the perspective of dynamical systems theories, the reinterpretation of the edge manifold $\Sigma$ as a boundary between two transition scenarios leads to interesting theoretical questions. One of the objectives of the description of the transition in terms of dynamical systems is to map out the state space of the system together with its invariant sets, at least in the context of an initial-value problem. Note that the Reynolds number is here not a parameter. The only fixed point of the system is the laminar state, which happens to be an unstable one, in the sense that there always exists perturbations likely to grow exponentially in time (without any parameter threshold to exceed). The turbulent attractor may not be within reach or even well-defined, but the edge state as relative attractor is well defined in the asymptotic limit $t \rightarrow +\infty$. In particular it is an invariant set for the temporal dynamics. It is also a self-similar object in terms of spatial dimensions and temporal dynamics. However the edge tracking technique, which by construction only identifies finite-time trajectories belonging to the stable manifold of the edge state, finds here a natural time limitation. This algorithm is based on bounds for a given observable $a$, originally designed to discard the basins of attraction of the other attracting sets. Here there is only one attracting turbulent set, therefore $\Sigma$ does not separate two distinct basins of attraction like in bistable systems. This is the main difficulty in interpreting phase portraits such as those in figure \ref{fig:phase_space}. In particular, no matter how small or how large the bound $a_L$, provided the computational box is long enough there will always be another trajectory with $a<a_L$ also experiencing transition, possibly featuring destabilisation of TS waves. A similar situation was described by \cite{zammert2017transition} in the context of plane Poiseuille flow in a small periodic domain, where the type of transition was determined from the knowledge of pre-defined transition times. In the present study, a suitable choice of the observable can extend the time over which the bisection stays valid. For instance, the use of the $\omega_x$ component to define the observable $a$ in equation \eqref{defa} is a convenient way to ignore Tollmien--Schlichting waves in their initial stage of their growth, because for TS waves it is non-zero only after they have undergone a secondary instability. Other one-dimensional observables can be envisioned. However without extensive monitoring of the properties of the trajectories generated during the bisection, a one-dimensional observable by itself does not contain enough information on the route to turbulence (bypass or classical) that is followed. Multi-dimensional observables, containing \textit{e.g.}\ information about amplitude or growth of the TS waves, are a possible alternative but they do not make bisection generically possible, since bisection is essentially a one-dimensional search process. For investigations where the asymptotic nature of the edge state matters, alternative methods are now welcome, specifically \emph{local} methods that do not require the knowledge of how trajectories behave far away in state space from the edge manifold under study.
\begin{figure}
\centering
  \includegraphics[width=12cm]{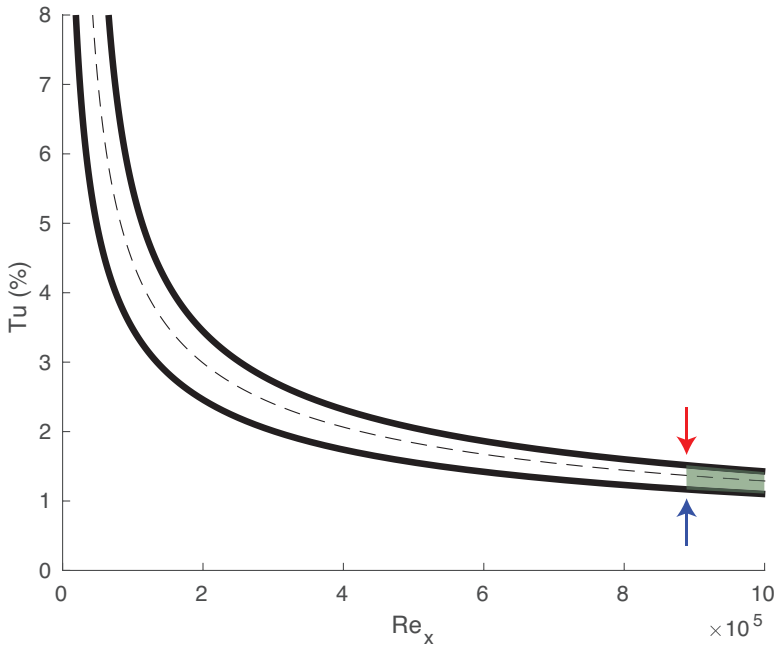}% Images in 100% size
  \caption{$Tu$($\%$) \textit{vs.}\ $Re_x$ for experimental bypass transition data \citep{ShahinfarFransson2011}, showing intermittency of 10\% and 90\% (thick lines) and 50\% (dashed line). The arrows mark the $Re_x$-limitation of the bisection algorithm. For larger times and $Re_x$, mixed transition is expected (grey area). The theoretical onset for TS waves lies at $Re_x=9.1\times 10^4$.} \label{fig:new_scenario}
\end{figure}

The possibility for proper coexistence of the two routes to turbulence sets a natural time limitation to the current algorithm of edge tracking. This time limit, estimated in Section 3 as about $4700 \delta_0^*/U_{\infty}$, is highlighted in figure \ref{fig:at}, which shows the time series of the observable $a(t)$ during the bisection process. It corresponds to a limitation of the maximum distance downstream along which edge tracking can be achieved, estimated here as $Re_x^{max} \approx 9 \times 10^5$. This is close to 10 times the value of $Re_x=9.1\times 10^4$ corresponding to the onset of TS waves according to linear stability analysis, and much larger than the value of $Re_x=3 \times 10^4$ corresponding to the inlet defined by $Re_{\delta^*_0}=300$. We now demonstrate that if the asymptotic edge state remains out of reach, the computed edge trajectories are relevant in practice for most instances of bypass transition under the influence of free-stream turbulence. We use for this purpose the experimental database by \cite{ShahinfarFransson2011} from the KTH wind tunnel with parametrisable free-stream turbulence, measured by $Tu=u_{rms}/U_{\infty}$ and expressed in $\%$. The $Re_x$-intervals where intermittency (and hence nucleation of turbulent spots) occurs are reported in figure \ref{fig:new_scenario} as functions of the parameter $Tu$, where the dependence on the integral length scale of the incoming turbulence is already taken into account.

The lower and upper bounds of these intervals are defined respectively by the values of $Re_x$ for which the intermittency factor is 0.1 and 0.9. They both scale like $Re_x=O(Tu^{-2})$ \citep{ShahinfarFransson2011}. For $Tu\approx2\%$ and above, the corresponding values of $Re_x$ in the diagram are always below the value of $Re_x^{max}$. This means that the edge state is known -and already computed- for these cases, and that it can be used as an alternative base flow for a stability analysis (generalised to unsteady flows). The only difficult situation corresponds to $Tu<2\%$, when the bisection technique is not able to track the edge all the way into the zone where spots nucleate. In this mixed area, both routes to turbulence are indistinguishable, at least from the monitoring of one scalar observable only. This mixed area is the parameter area where the turbulent patches observed in practice can be due to the destabilisation of either TS waves or streaky perturbations. It has long been known as a delicate range of parameters and has sometimes been labelled ``weak bypass regime'', as opposed to the ``strong bypass regime'' where streak breakdown is the sole cause for transition \citep{Narasimha1994}. Future efforts should be made to better characterise the complex transition to turbulence in this intermediate regime. 

B. Eckhardt from Philipps University Marburg is acknowledged for interesting discussions. Financial support by the Swedish Research Council (VR) grant no.~2016-03541 is gratefully acknowledged. The computations were performed on resources provided by the Swedish National Infrastructure for Computing (SNIC) at PDC and NSC. 

\appendix
\section{}\label{appA}
%\highlight{MB: Under construction, which figures should we keep here? How much should we write about the limitations of the moving box and the waves coming in?}

In order to implement the moving box it is necessary to do a Galilean change of reference and to update the base flow enforcing correct inflow and outflow conditions while preserving streamwise periodicity. The update of the base flow can be explained as follows. We first consider the incompressible Navier--Stokes equations:
\begin{align}
   % \frac{\partial u_i}{\partial t}&=-\frac{\partial p}{\partial x_i}+\epsilon_{ijk}u_j\omega_k-\frac{\partial}{\partial x_i}\left(\frac{1}{2}u_j u_j \right)+\frac{1}{Re}\nabla^2u_i+F_i, \\
    \partial_t v_i + v_j \partial_j v_i &=-\partial_i p + Re_{\delta_0^*}^{-1}\partial^2_j v_i + F_i, \\
    \partial_j v_j=0.    
\end{align}

%\begin{align}
%   % \frac{\partial u_i}{\partial t}&=-\frac{\partial p}{\partial x_i}+\epsilon_{ijk}u_j\omega_k-\frac{\partial}{\partial x_i}\left(\frac{1}{2}u_j u_j \right)+\frac{1}{Re}\nabla^2u_i+F_i, \\
 %   \frac{\partial \textbf{v}}{\partial t} + \left( \textbf{v} \cdot {\bm \nabla}\right) \textbf{v} &=-{\bm \nabla} p + \frac{1}{Re_{\delta_0^*}}{\bm \nabla}^2\textbf{v}+\textbf{F}, \\
%    {\bm \nabla} \cdot \textbf{v}&=0.    
%\end{align}

The perturbation velocity field $\textbf{u}$ is deduced from the full velocity field $\textbf{v}$ by substracting the analytical Blasius profile $\textbf{v}^B=(u_x^{B},v_x^{B},0)$, itself an excellent approximation to the steady base flow for $x \gg 1$ \citep{schlichting2016boundary}. The field $\textbf{F}=(F_x,F_y,F_z)$ is a volume force imposed in the fringe region in order to ensure the $x$-periodicity in spatial simulations. The form of the forcing is:
\begin{equation}
    \textbf{F} = \gamma(x)(\bm{\mathcal{U}}(t)-\textbf{v}),
\end{equation}
where $\textbf{v}$ is the instantaneous flow field and $\gamma(x)$ is a non-negative fringe function. The function $\gamma(x)$ is non-zero in the fringe region only. The streamwise component of $\bm{\mathcal{U}}(t)$ is computed as \citep[see][]{chevalier2007simson}
\begin{equation}
    \mathcal{U}(x,y,z,t) = U(x,y,z,t)+\left[ U(x+x_L,y,z,t)-U(x,y,z,t)\right]S\left(\frac{x-x_{\text{blend}}}{\Delta_{\text{blend}}}\right), 
\end{equation}
where $S(x_{\text{blend}},\Delta_{\text{blend}})$ is a step function, and where $x_{\text{blend}}$ and $\Delta_{\text{blend}}$ are chosen so that the flow connects smoothly the outflow to the inflow. The normal component of $\bm{\mathcal{U}}(t)$ is computed from continuity. The novelty in our approach is that $U(x,y,z,t)$, the solution to the boundary layer equations, is allowed to change with time, thereby modifying the base flow on which the fringe acts. $U(x,y,z,t)=v_x^{B}(x,y,x_0(t))$ is updated at time $t$, such that the inlet of the box $x_0$ for which $\delta_0^*=1$ at $Re_{\delta^*}=300$ reads $x_0 \leftarrow x_{0}+c_{\text{box}}t_{m}$, where $t_{m}$ is the total time during which the box has moved with velocity $c_{\text{box}}$.

\bibliographystyle{jfm}
%\bibliography{bibliography_phd.bib}
\bibliography{2018_edgeBL_JFM.bib}

\end{document}